\documentclass[11pt]{article}

\usepackage{authblk}
\usepackage{mathtools}
\usepackage{amsmath}
\usepackage{amsfonts}
\usepackage{amssymb}
\usepackage{listings}
\usepackage{graphicx}
\usepackage{longtable}
\usepackage[parfill]{parskip}
\usepackage{bbm}
\usepackage{float}
\usepackage{adjustbox}
\usepackage{algorithm}
\usepackage{algorithmic}
\usepackage{caption}
\usepackage[caption=false]{subfig}
\usepackage{hyperref}
\usepackage{multirow}
\usepackage{rotating}
\usepackage{newtxtext, makeidx, multicol, footmisc}
\usepackage[shortlabels]{enumitem}

\usepackage{xcolor}
\usepackage{float}

\usepackage{lineno}
\title{A statistical test for network similarity}

\author[1]{Pierre Miasnikof \thanks{corresponding author: pierre.miasnikof@fsa.ulaval.ca}}
\author[2,3]{Alexander Y. Shestopaloff}

\affil[1]{Université Laval, Québec, QC, Canada}
\affil[2]{Queen Mary University of London, London, United Kingdom}
\affil[3]{Memorial University of Newfoundland, Saint-John's, NL, Canada}

\date{}
\begin{document}
\maketitle
	
\begin{abstract}
In this article, we revisit and expand our prior work on graph similarity. As with our earlier work, we focus on a view of similarity which does not require node correspondence between graphs under comparison. Our work is suited to the temporal study of networks, change-point and anomaly detection and simple comparisons of static graphs. It provides a similarity metric for the study of (weakly) connected graphs. Our work proposes a metric designed to compare networks and assess the (dis)similarity between them. For example, given three different graphs with possibly different numbers of nodes, $G_1$, $G_2$ and $G_3$, we aim to answer two questions: a) "How different is $G_1 $ from $G_2$?" and b) "Is graph $G_3$ more similar to $G_1$ or to $G_2$?". We illustrate the value of our test and its accuracy through several new experiments, using synthetic and real-world graphs.
\end{abstract}

\section{Introduction}
In this article, we revisit and expand our prior work on graph similarity \cite{StatSim2023}. As with our earlier work, we focus on a view of similarity which does not require node correspondence between graphs under comparison. Our work is suited to the temporal study of networks, change-point and anomaly detection and simple comparisons of static graphs. Naturally, change-point or anomaly detection can be performed with our test, but it should be coupled with area expertise and a thorough examination of each graph's structure. Our work provides a similarity metric. However, the significance of this (dis)similarity is more a structural and domain question than a statistical one.

While our earlier work did not explicitly mention this applicability condition of (weak) connectedness, we wish to clearly state it here. In this updated version, we show extensive empirical testing and a detailed sensitivity analysis. In addition, we also extend our test to (weakly connected) directed graphs. Our work proposes a metric designed to compare networks and assess the (dis)similarity between them. For example, given three different graphs with possibly different numbers of nodes, $G_1$, $G_2$ and $G_3$, we aim to answer two questions: a) "How different is $G_1 $ from $G_2$?" and b) "Is graph $G_3$ more similar to $G_1$ or to $G_2$?".

To this day, graph isomorphism remains a challenging problem, with no known polynomial-time solution. Meanwhile, the broader tasks of calculating graph similarity metrics, such as graph edit distance or maximum common subgraph, are NP-hard. However, with the growing availability of graph datasets, from social networks and communication networks to financial systems, network comparison and analysis have become crucial \cite{Tan2019,Wills2020,Piccardi2025}.
	
To meet this need to compare graphs, several methods have been created \cite{Soundar2013}. Our method stands out from others because of its statistical grounding, its integration of node-level, neighborhood-level and graph-wide features and polynomial time complexity.
	
We begin by converting networks (graphs) into an all pairs distances matrix. We choose the Jaccard distance instead of the typical shortest-path or random walk-based distances (e.g., commute, resistance,...). This choice is not arbitrary, it is justified by work that has identified the limitations of shortest-path \cite{RDist2021} and random walk-based distances \cite{vonLuxNIPS10,vonLux14,medoidsPPS19,StatSim2023}. In contrast, Jaccard distance has been shown to capture the graph's structural characteristics \cite{Camby17,PMCplxNets2020,PMCplxNets2022,StatSim2023}.
	
Our comparison begins with node and neighborhood-level characteristics, but aggregates these at a network-level, in order to offer graph-level metrics. We argue this transformation provides a more informative viewpoint on the entire network than easily observable graph characteristics, such as the number of nodes or edges and even density and degree distribution. We posit that connectivity and changes in connectivity represent the graph's very definition, structure and evolution. 
	
For example, in a physical or computer network, the emergence of denser subgraphs may indicate a loss of connection to the broader network or the appearance of bottlenecks. They can also be an indicator of malicious activity, especially of the multi-party coordinated variety \cite{rla2008,rla2011,denseAnom2021}.  
	
Above all, Jaccard distance can be interpreted through a probabilistic lens. It is because of this interpretation that we are then able to compare networks as probability distributions of node-node distances, by applying common statistical techniques. This conversion to a distance matrix and its probabilistic examination as a distribution offers a broader structural viewpoint of the graph than the focused examination of specific graph attributes.
	
In summary, this work offers a probabilistic structural viewpoint as basis for comparing graphs, without imposing restrictive theoretical models. Our work is suited to the temporal study of networks, change-point and anomaly detection and simple comparisons of static graphs. To our knowledge, no other technique in the literature offers such a comprehensive and easily interpretable comparison of graphs.
	
\section{Previous work}
In our discussion of the work done to this point, we would like to begin with a justification for what we deliberately omitted, a discussion of graphons. While we are aware of this very active area of the literature, we do not consider it well-suited to the objectives of our research. Nevertheless, we feel it is necessary to reassure readers that graphons are not competitors for the solution we propose in this work. 

Graph comparisons can certainly be done through the use of (empirical) graphons (e.g., \cite{Lovasz2012}). Unfortunately, comparison via graphons can be computationally expensive, especially when the alignment step requires solving an optimization problem over all vertex permutations. Because of the unsuitability of graphons to sparse networks and because of the depth of the non-graphon graph comparison literature, graphons are not considered in this work.
	
The study of network similarity is well-established and wide-ranging (e.g., \cite{Sanfeliu83,Bunke2000,Berling2012,Soundar2013,Tan2019,Piccardi2025}). Case in point, Soundarajan et al.~reported in 2013 that there were several graph similarity assessment techniques and that most offered very similar results. These authors also offered a classification of network similarity techniques. They observed that most of these algorithms analyzed either node-level, community-level or network-level data. Interestingly, on the topic of data models, only a few months earlier, Berlingerio et al.~(2012) \cite{Berling2012} highlighted the need to encode and summarize graphs into a ``signature vector''. As stated earlier, we feel that our proposed technique transcends these data partitions and offers a three-level point-of-view.
	
To this day, network similarity remains an active area of research (e.g. \cite{Piccardi2025}). For this reason, a complete review of the literature is a task in itself, a task well beyond the scope of this article. Therefore, in this section we focus on the work we use as the building blocks of our own work.
	
Our work is heavily influenced by that of Schieber et al.~(2017) \cite{SchieberEtAl2017}. Following their example, we study the distribution of node-node distances. However, while these authors study the distribution of shortest-path distance between nodes, we use Jaccard. We are also guided by the work of Wang et al.~(2022) \cite{WangEtAl2022}. Like Wang et al., we also examine the distribution of node to node distance, but with more clearly obtained distances. In contrast to them, we do not compute distances between vector representations of nodes that are obtained by random walk distances \cite{DeepWalk2014}. Unlike the work described in these two publications, in light of the overhead imposed by random walk simulations, the limitations of random walk \cite{vonLuxNIPS10,vonLux14,medoidsPPS19,StatSim2023} or shortest path distances \cite{ChebotarevShamis2006a,FoussEtAl2012,RDist2021} and the work of Camby and Caporossi (2017) \cite{Camby17}, we use Jaccard distance between nodes. In fact, the details and benefits of using Jaccard for node similarity have been well-documented \cite{JaccOrig,Camby17,PMCplxNets2020,PMCplxNets2022}.
	
We also borrow the idea of conducting statistical comparisons of graphs from Peel and Clauset (2015) \cite{Peel2015}. However, while these authors do conduct change-point detection using statistical techniques, they begin by imposing a statistical model on their graph under study and then conduct tests within its constraints. In contrast, we do not assume any distribution for our graph data.
	
Finally, we would like to end this review with a word on comparisons based on graph spectra (e.g., \cite{HuangEtAl2020}). In addition to their computational overhead, such techniques do not easily extend to directed graphs. Furthermore, graph spectra have been found to be very sensitive to graph size and noise in connectivity patterns and to offer somewhat unreliable pictures of the graphs they represent \cite{SpectralLim2024}.
	
\section{Methods}
Our (dis)similarity algorithm involves three distinct steps. First, we convert each (weakly) connected graph into an all-pairs Jaccard distance. We then interpret these distances  matrices as sets of random variables. Finally, we measure the (dis)similarity between each graph (random variable set) using the Kolmogorov-Smirnov metric \cite{KSTest,WasKS2017}.
	
\subsection{All-pairs distance}
As described in our earlier work \cite{StatSim2023}, our analysis begins with a transformation of each graph being compared into an all-pairs Jaccard distance matrix. For each pair of vertices in a given graph we compute the following distances. 
	
In the general (and undirected) case, the Jaccard distance separating two vertices $i$ and $j$ is defined as
\[
	\zeta_{ij} = 1 - \underbrace{\frac{ \vert a_i \cap a_j \vert  }{ \vert a_i \cup a_j \vert}}_{\sigma_{ij}} \in [0,1] \, .
\]
These distances are then recorded into a $\vert V \vert \times \vert V \vert$ hollow and symmetric matrix $Z = \left[ \zeta_{ij} \right]$, where $\zeta_{ij} = \zeta_{ji}$ and $\zeta_{ii} = 0, \, \forall i,j$. Here, $a_i \, (a_j)$ represents the set of all vertices with which vertex $i \, (j)$ shares an edge. The ratio $\sigma_{ij}$ is the well known Jaccard similarity. The Jaccard distance ($\zeta_{ij}$) is its complement.
	
As mentioned earlier, in this version of our work, we extend this distance to directed graphs. In the case of directed graphs, we simply decompose the cardinalities of the intersection and union of shared connections into the sum of the cardinality of their two constituents:
	\[
	\zeta_{ij} = 1 - \underbrace{\frac{ \vert p_i \cap p_j \vert +  \vert \sigma_i \cap \sigma_j \vert}{\vert p_i \cup p_j \vert +  \vert \sigma_i \cup \sigma_j \vert}}_{s_{ij}} \in [0,1] \,. 
	\]
	In this specific case, $p_i \, (p_j)$ is the set of predecessor nodes to node $i \, (j)$. Meanwhile, $\sigma_i \, (\sigma_j)$ is the set of successor nodes.

\subsubsection{Probabilistic interpretation of the Jaccard distance}
As outlined in our previous work, Jaccard similarity ($s_{ij}$) between two nodes $i$ and $j$ has a probabilistic interpretation, in the undirected case \cite{StatSim2023}. Indeed, Jaccard similarity is an estimate of the conditional probability that both $i$ and $j$ are connected to $k$, given that at least one of $i$ or $j$ is connected to $k$. Mathematically, we expressed $s_{ij}$ as
	\[
	s_{ij} = P \left( \left( e_{ik} \land e_{jk} \right) \mid \left( e_{ik} \lor e_{jk)} \right) \right) \, ,
	\] where $e_{ij}$ indicates the existence of an edge between nodes $i$ and $j$.
	
Naturally, Jaccard distance is its complement (i.e., $\zeta_{ij} = 1 - s_{ij}$). Therefore, it can be interpreted as one of these two cases,
	\begin{enumerate}[a)]
		\item the conditional probability that $i$  is connected to $k$, but $j$ is not,
	\end{enumerate}
	(exclusive) or
	\begin{enumerate}[b)]
		\item  the conditional probability that $j$  is connected to $k$, but $i$ is not.
	\end{enumerate}
Mathematically, we express Jaccard distance as
	\[
	\zeta_{ij} = 1 - s_{ij} = 1 - P \left( \left( e_{ik} \land e_{jk} \right) \mid \left( e_{ik} \lor e_{jk)} \right) \right) = P \left( \left( e_{ik} \veebar e_{jk} \right) \mid \left( e_{ik} \lor e_{jk)} \right) \right) \, . 
	\]
In the specific case of directed graphs, the $\zeta_{ij}$ has an essentially similar interpretation, but with edge direction ignored.
	
\subsubsection{From graph to empirical probability distribution} 
With all distances $\zeta_{ij}$ computed, we then take a graph-wide point of view. In light of the probabilistic interpretation of the $\zeta_{ij}$ described previously, we  compare graphs as empirical probability distributions of node-node distances. In other words, each node-node distance is understood to be a random variable and every graph is examined as an empirical statistical distribution.
	
\begin{figure}[]
		\centering
		\subfloat[ER]{ \includegraphics[width = 0.48\textwidth]{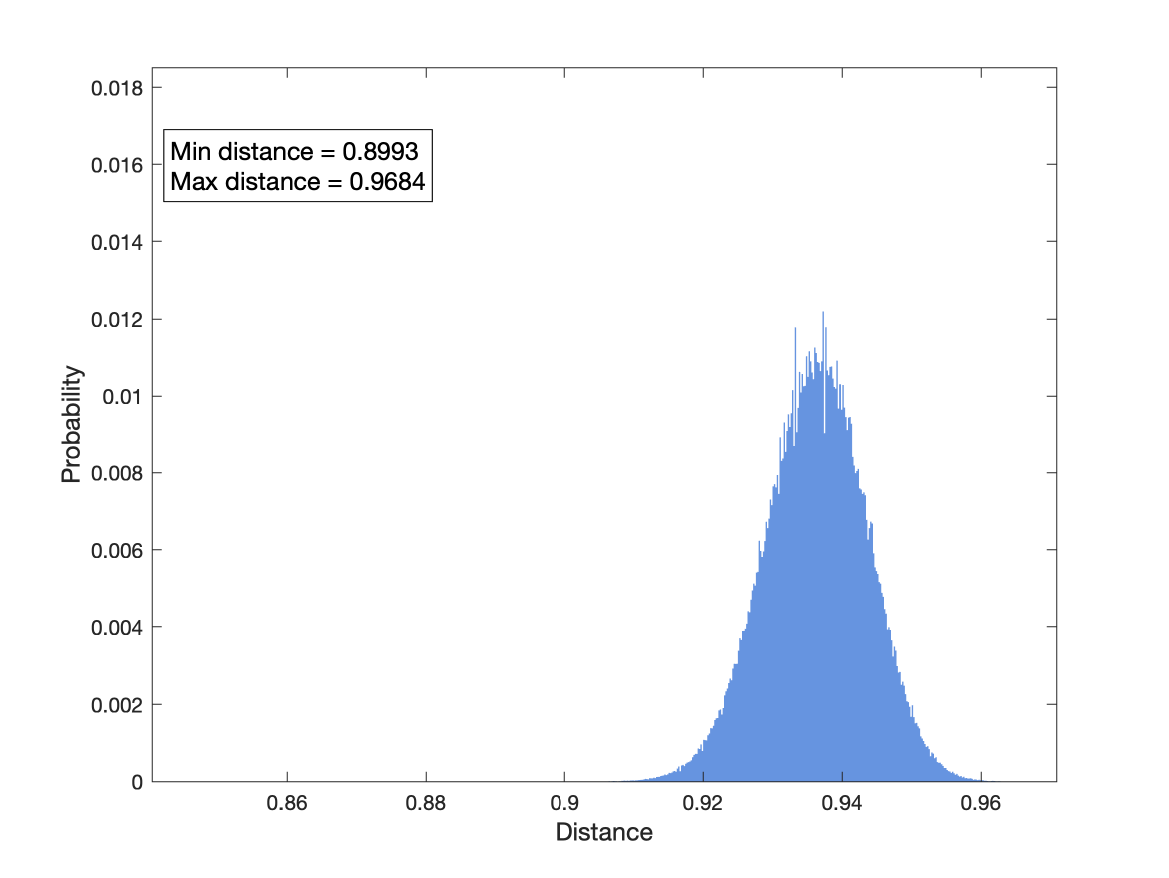} } 
		\subfloat[PPM]{ \includegraphics[width = 0.48\textwidth]{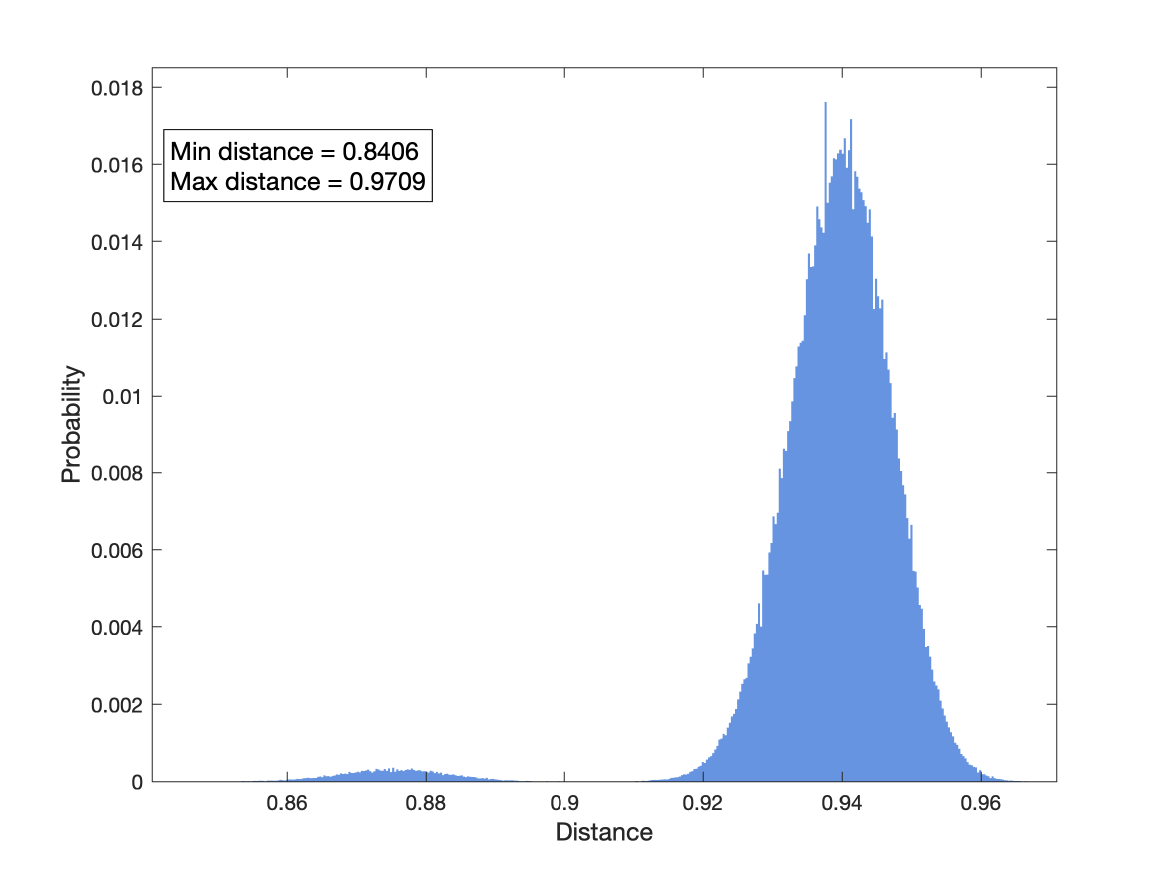} } 
		\caption{Distances as distributions. The distribution of node-to-node distances for the Erd\H{o}s-R\'enyi (ER) graph is unimodal while it is bimodal for the planted partition model (PPM) graph.}
		\label{distn}
	\end{figure}

Figure~\ref{distn}, borrowed from our previous work on the subject \cite{StatSim2023}, provides an example of the application of the core ideas of this work. On the left, we see the distribution of node-node distances for an  Erd\H{o}s-R\'enyi (ER) graph \cite{ER1959,Gilbert1959} with 5,000 nodes and an edge probability of $p=0.5$. On the right, we see the distribution of distances between nodes of a planted partition model graph (PPM) \cite{PPM2001} with 50 blocks of 100 nodes and in/out edge probabilities of $0.9/0.1$. (Both graphs have the same number of nodes, both distributions contain the same number of observations.)

Here, it is clear that the very significant structural difference between these two graphs is very well translated into the difference between the distributions of node-node distances. The ER graph's distances are symmetrically distributed about their mean, in a Gaussian-like pattern. Markedly, the PPM graph's distances are left-skewed and bi-modal. In this specific insttance, the left mode reflects distances between nodes in the same blocks, whereas the right mode reflects distances between nodes not in the same blocks. Naturally, this bi-modal pattern does not appear in the case of the ER graph.
	
\subsection{(Dis)similarity of probability distributions}
With each network's empirical probability distributions of the Jaccard distances between their nodes, we compute the Kolmogorov-Smirnov (K-S) distances separating them \cite{KSTest}. In our previous work, we also examined the Wasserstein distances \cite{WasKS2017}. However, due to the extremely consistent results we observed between the two techniques and the ease of interpretation of the K-S distances, we only use K-S in this updated analysis. (The consistency between K-S and Wasserstein was also documented by Ramdas et al.~\cite{WasKS2017}.)

When comparing two networks, we define the K-S distance as follows:
\begin{itemize}
	\item Let $F_{1}(x)$ be the empirical cumulative distribution function (CDF) of Jaccard distances for the first network
	\item Let $F_{2}(x)$ be the empirical CDF of Jaccard distances for the second network 
	\item The Kolmogorov-Smirnov (K-S) distance to compare $F_{1}$ and $F_{2}$ is defined as
		\begin{equation*}
		D = \sup_{x} \vert F_{1}(x) - F_{2}(x) \vert \quad (\in [0,1]) \, .
		\end{equation*}
\end{itemize}
	
The K-S distance metric $D$ is also a test statistic. In this specific case, it is a test statistic for the two-sample K-S test. The hypotheses of this test are listed below.
	\begin{itemize}
		\item Null hypothesis ($H_o$) : the two samples are drawn from the same distribution
		\item Alternative hypothesis ($H_a$) : the two samples are drawn from different distributions
	\end{itemize}
The $p$-values of the K-S test provide an interpretation and validation of the test statistic (distance $D$). Concretely, they are the measure of the area under the Kolmogorov distribution's probability density curve beyond the point $D$. Ideally, this area represents the probability of obtaining a distance of the same or greater magnitude, under the (null) hypothesis that both samples were drawn from the same distribution. Under theoretical circumstances, small $p$-values provide evidence that the maximum vertical distance between the empirical CDFs of two compared graphs is statistically significantly different from zero. 

Unfortunately, the K-S test requires that both the observations within each distribution and the distributions being compared themselves be independent and identically distributed (iid). While the $p$-values in our test do behave exactly as expected, they cannot be relied on for their statistical accuracy. For this reason, we focus our attention on the K-S distances themselves. After all, even with moderately sized graphs like the one in our experiments (e.g., $N = 5,000 \Rightarrow \sim 5,000^2$ distances), the sizes of our samples are extremely large. Argubaly, these extremely large samples make our distances very reliable metrics of the (dis)similarity between the distributions being compared. For this reason, under our simplified test, we consider two graph to be dissimilar, when the K-S distance is greater than zero, without regard for the significance of this figure.

\section{Numerical results}
We conduct four types of numerical experiments. In Section~\ref{2graphtests}, we compare graphs that are known to be different. The goal of these comparisons is to illustrate our test. These comparisons are new instances of the tests we conducted in our previous work \cite{StatSim2023}. 

In Section~\ref{sensitivity}, we compare individual graphs with versions of themselves with randomly selected nodes or edges having been removed. The goal of these experiments is, in part, to validate the correct verdict of similarity, in cases where we compare the graph to an unaltered version of itself. These trials also offer insight into the sensitivity of our test to graph structure modification through node or edge removal (e.g., in cases of temporal or change-point detection settings). 

It is important to note here that these modifications to graphs (node or edge removal) may or may not constitute a significant or even observable modification to the initial graph. The experiments in this section consist of the same modification to graphs with varied structures. The goal of these experiments is not to set a threshold for two graphs to be declared ``significantly dissimilar''. Rather, the goal is to examine if and how these modifications will be reflected in the K-S distances between the Jaccard distance distributions. Naturally, change-point or anomaly detection can be performed with our test, but it should be coupled with area expertise and a thorough examination of each graph's structure.

Finally, our fourth set of experiments, shown in Section~\ref{adds}, is a comparison between graphs with an added number of nodes and randomly generated connections between them and their initial graph version. These experiments validate the applicability of our tests and its correct conclusions (similar or not), in the case of the very sparse graphs that were unsuited to our edge and node removal experiments in Section~\ref{sensitivity}. Here again, these experiments aim to test the sensitivity or our metric. They consist of the same modification to graphs with varied structures. These modifications may or may not be significant, given each initial graph's structure.

\subsection{Two-graph tests} \label{2graphtests}
\begin{figure}[H]
  \centering
  \subfloat[ER\;0.309]{%
    \includegraphics[width=0.48\textwidth]{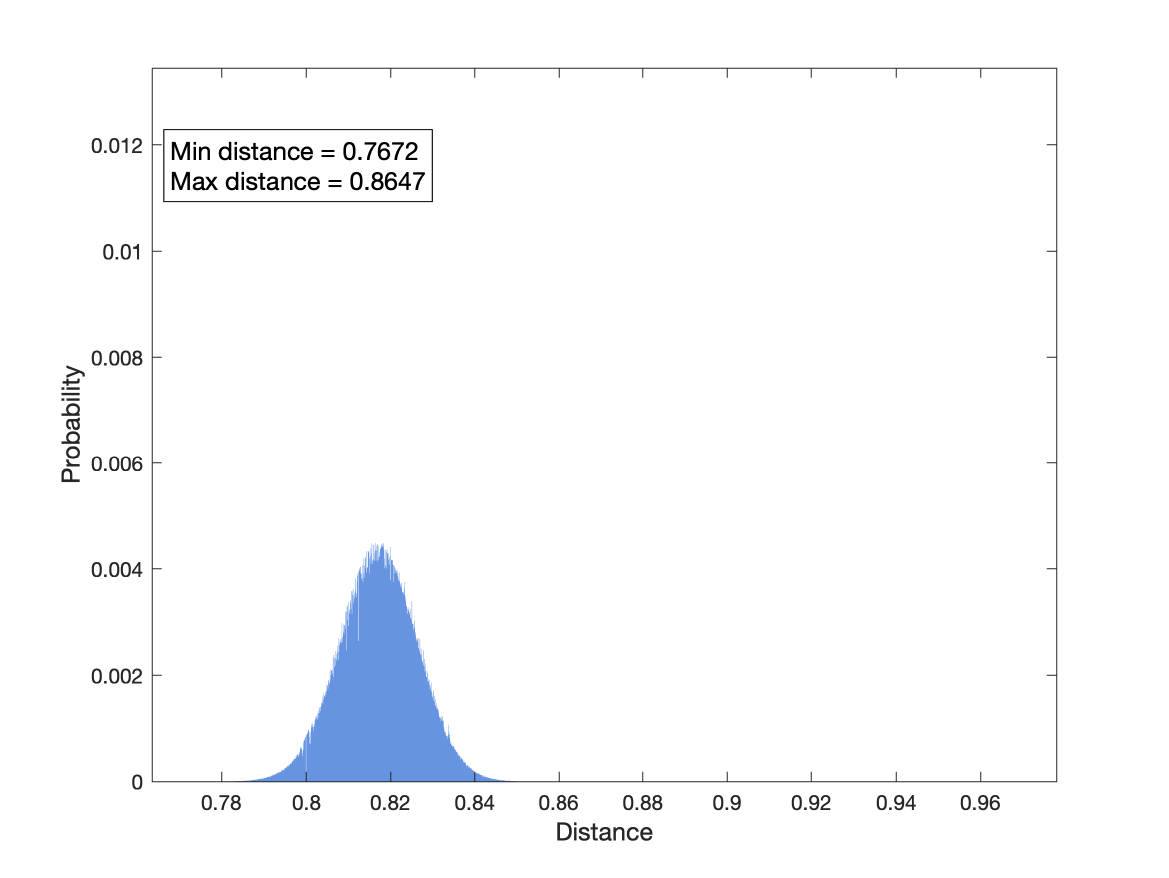}
  }\hfill
  \subfloat[ER\ (Directed)\;0.309]{%
    \includegraphics[width=0.48\textwidth]{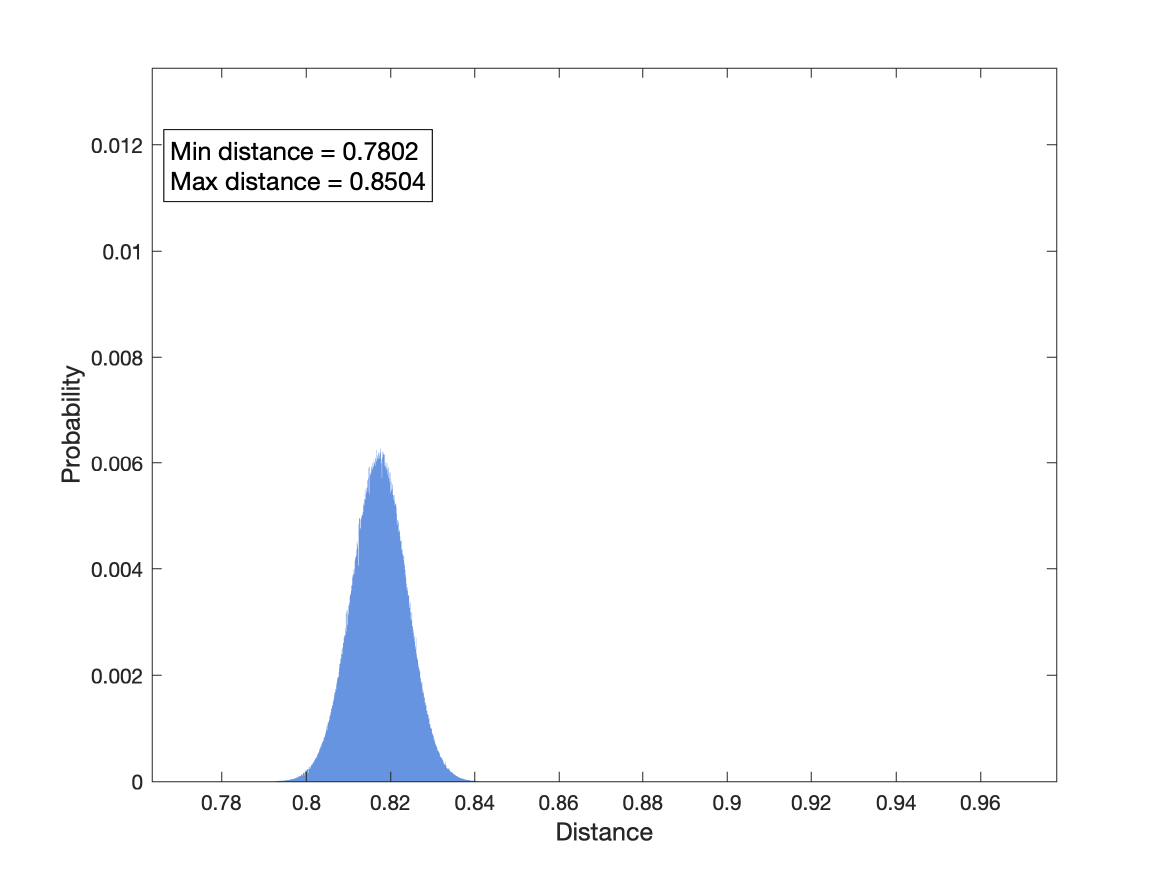}
  }\\[0.75ex]
  \subfloat[ER\;0.213]{%
    \includegraphics[width=0.48\textwidth]{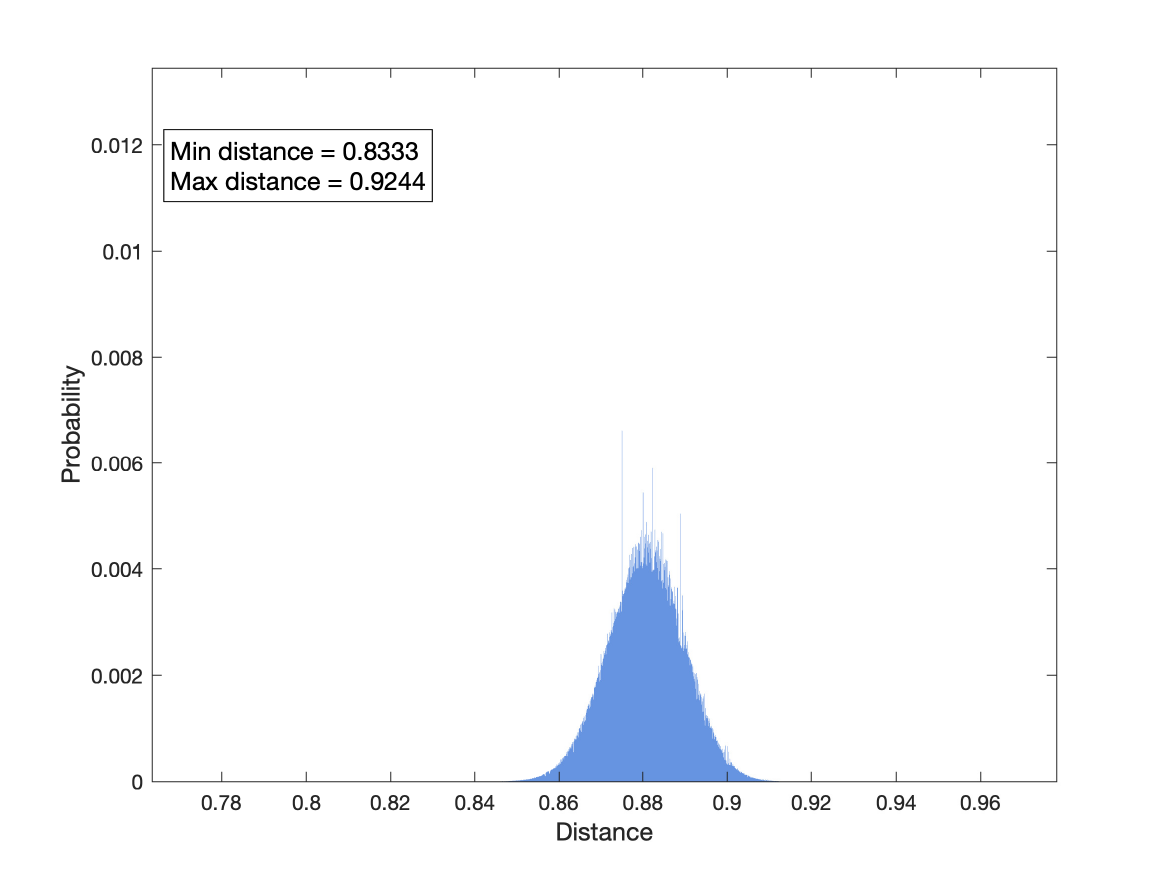}
  }\hfill
  \subfloat[ER\ (Directed)\;0.213]{%
    \includegraphics[width=0.48\textwidth]{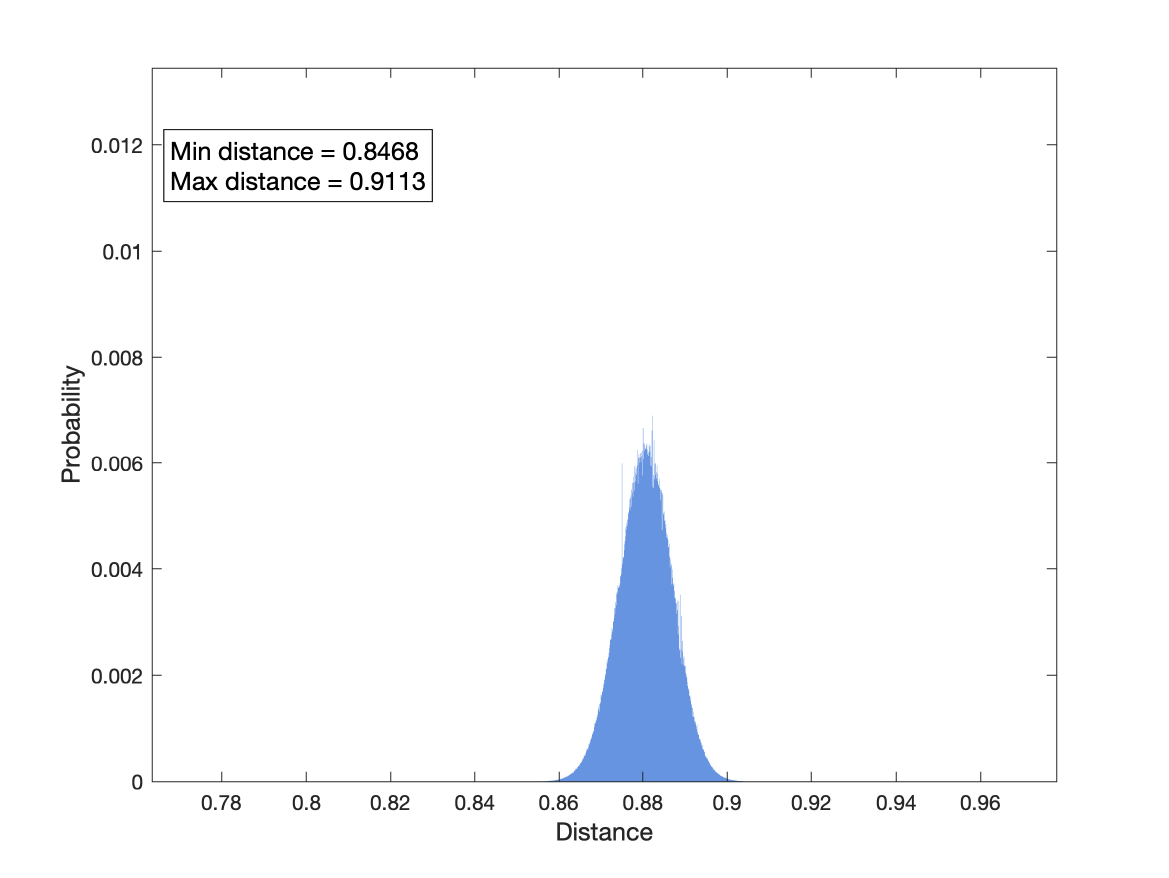}
  }\\[0.75ex]
  \subfloat[ER\;0.118]{%
    \includegraphics[width=0.48\textwidth]{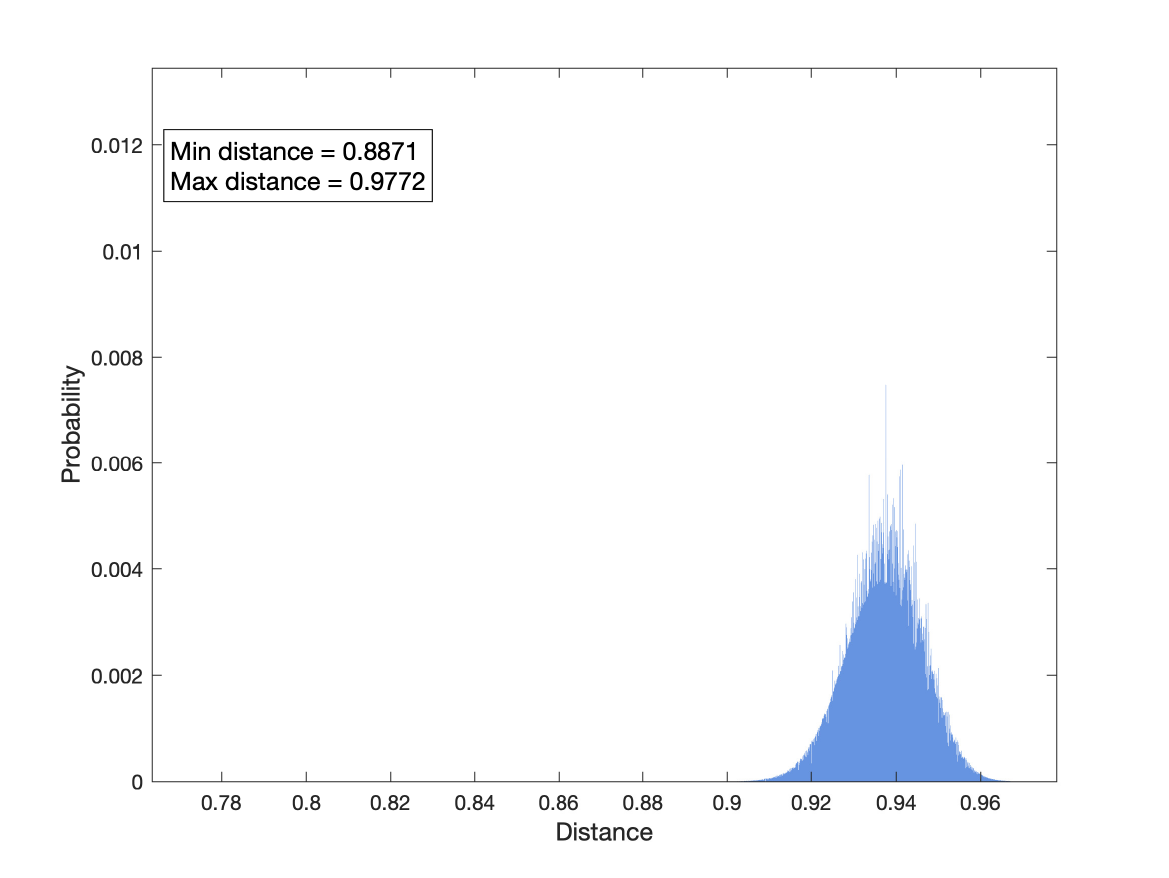}
  }\hfill
  \subfloat[ER\ (Directed)\;0.118]{%
    \includegraphics[width=0.48\textwidth]{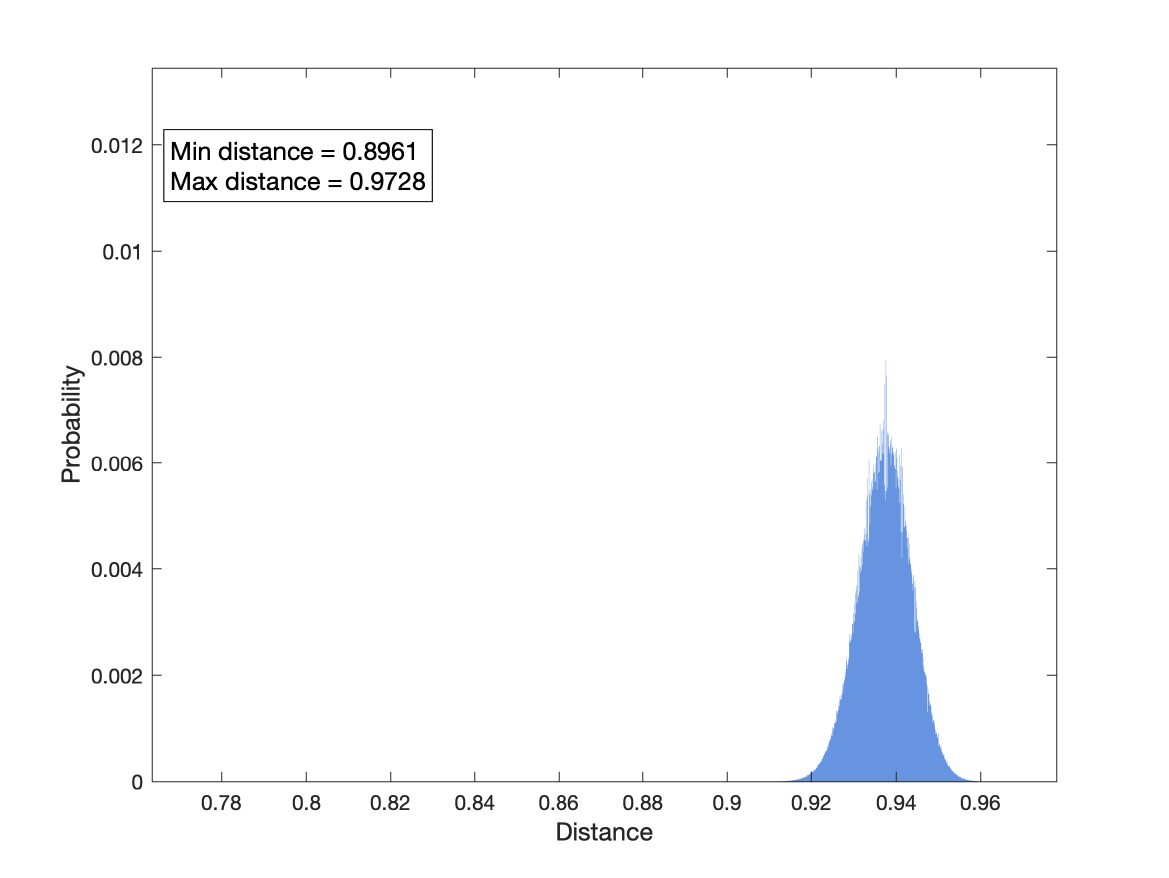}
  }
  \caption{Jaccard distance distributions for ER models with given densities.}
  \label{fig:er-graphs}
\end{figure}

\begin{figure}[H]
  \centering
  \subfloat[SBM\;0.7/0.3]{%
    \includegraphics[width=0.48\textwidth]{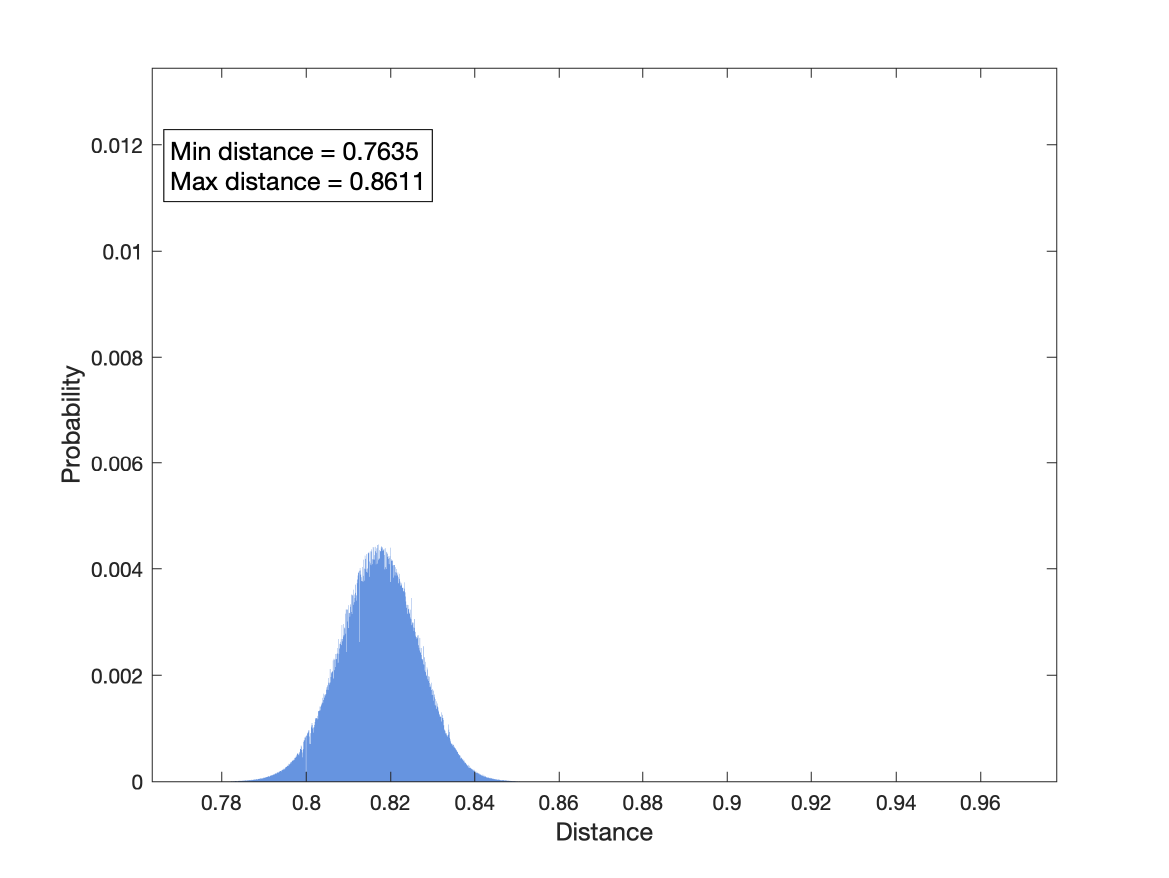}
  }\hfill
  \subfloat[SBM\ (Directed)\;0.7/0.3]{%
    \includegraphics[width=0.48\textwidth]{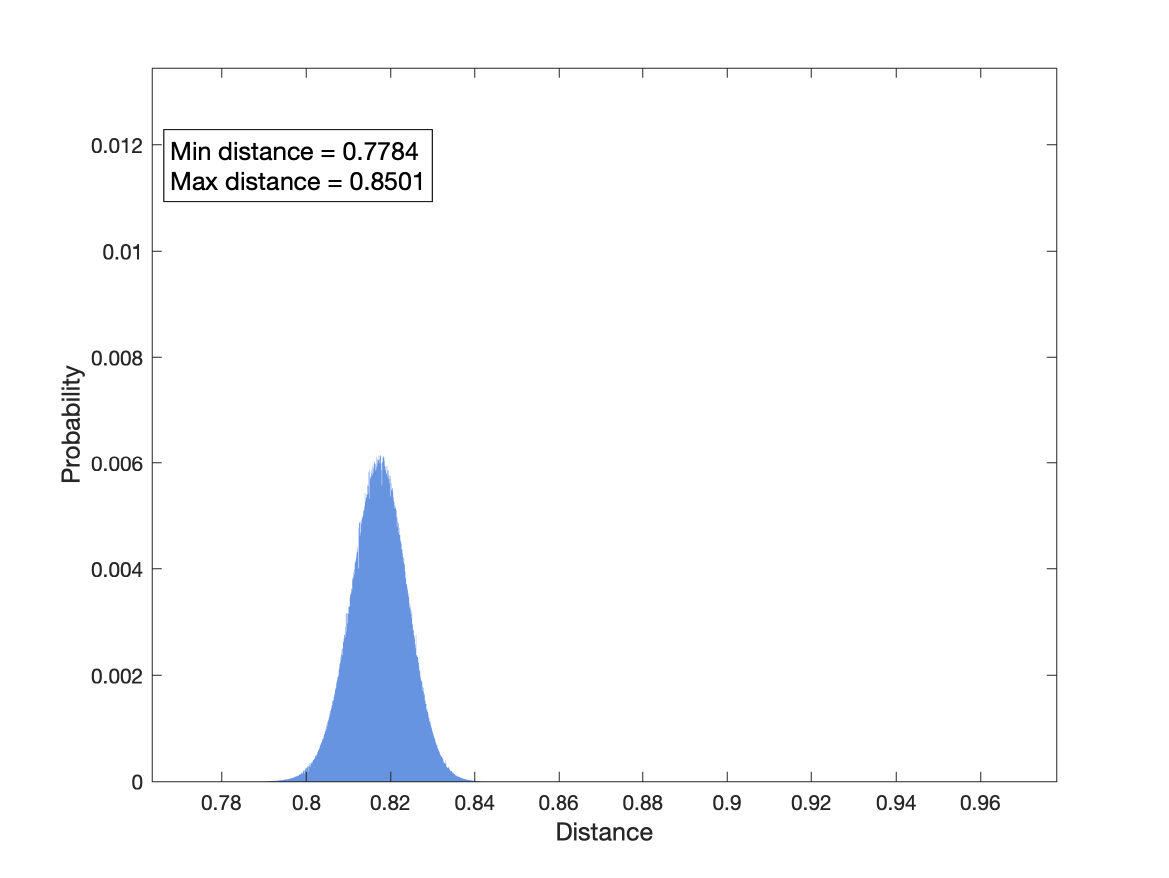}
  }\\[0.75ex]
  \subfloat[SBM\;0.8/0.2]{%
    \includegraphics[width=0.48\textwidth]{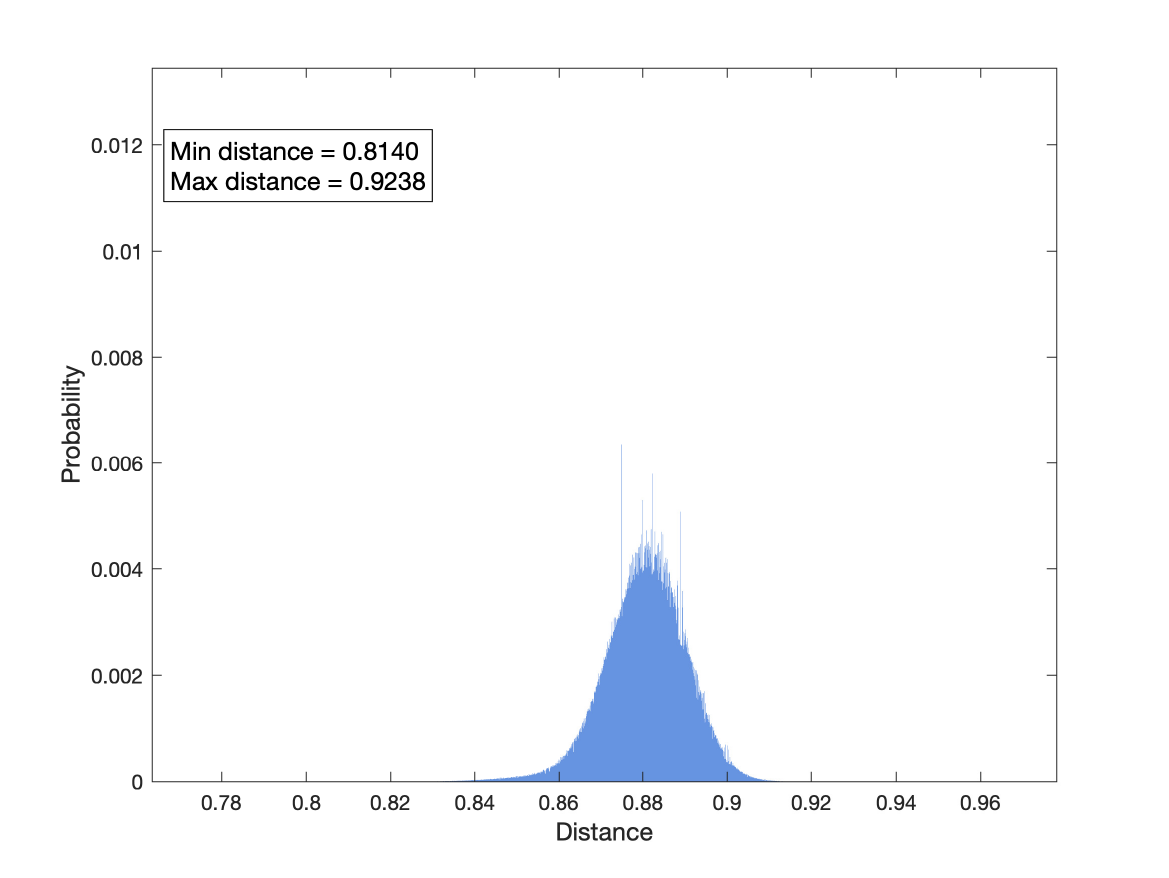}
  }\hfill
  \subfloat[SBM\ (Directed)\;0.8/0.2]{%
    \includegraphics[width=0.48\textwidth]{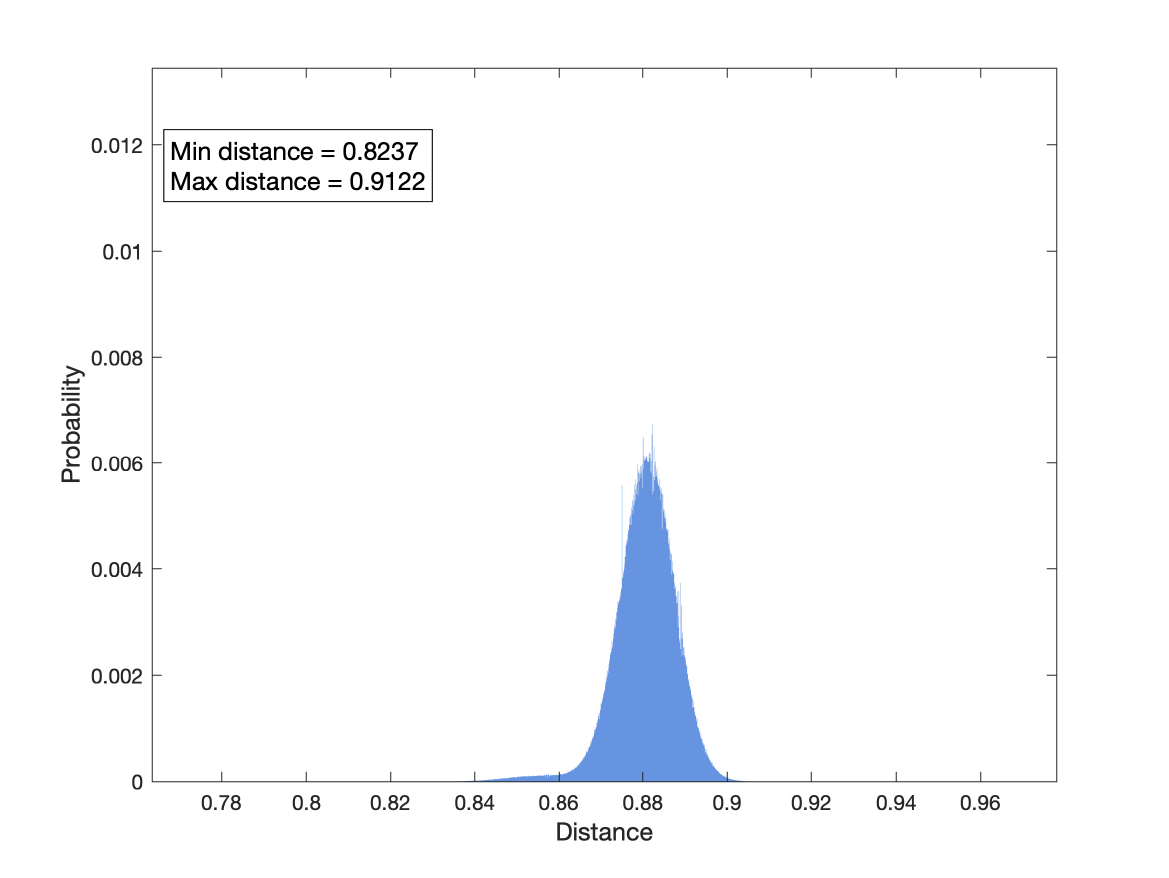}
  }\\[0.75ex]
  \subfloat[SBM\;0.9/0.1]{%
    \includegraphics[width=0.48\textwidth]{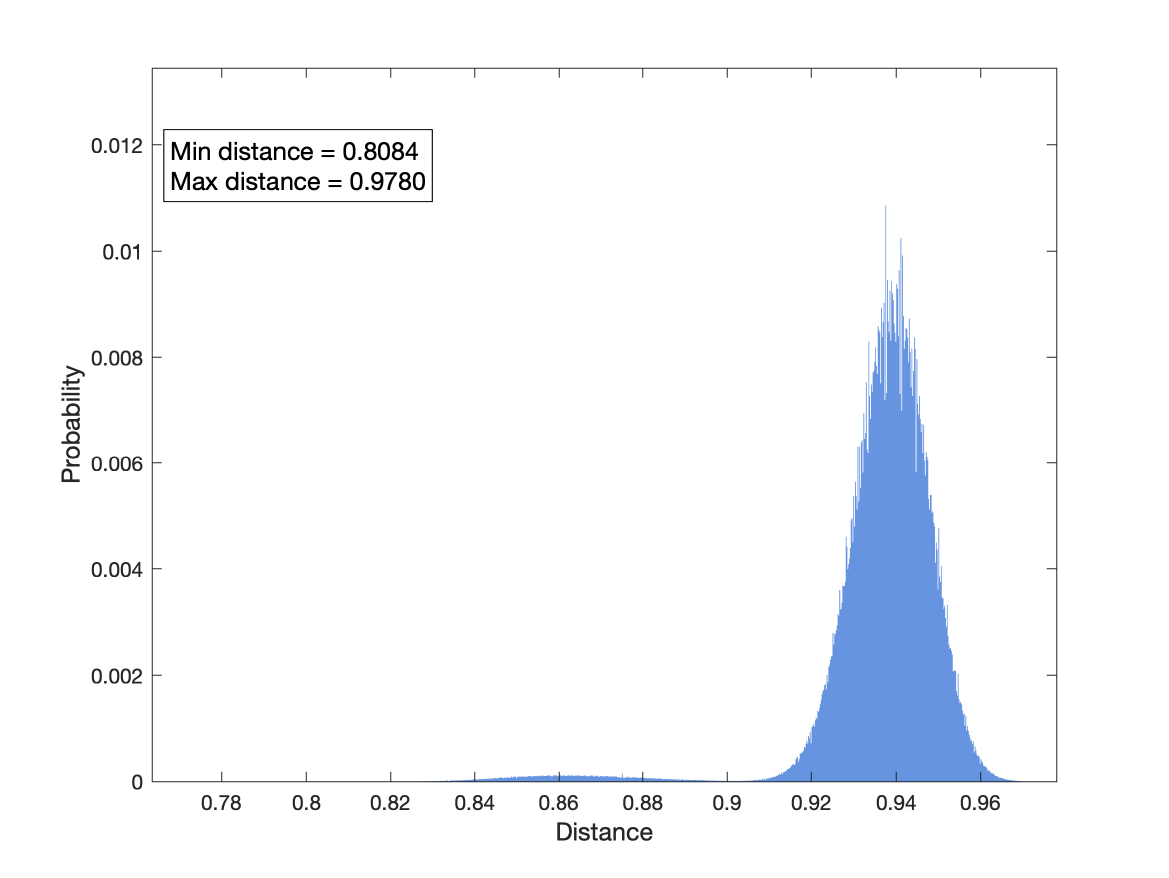}
  }\hfill
  \subfloat[SBM\ (Directed)\;0.9/0.1]{%
    \includegraphics[width=0.48\textwidth]{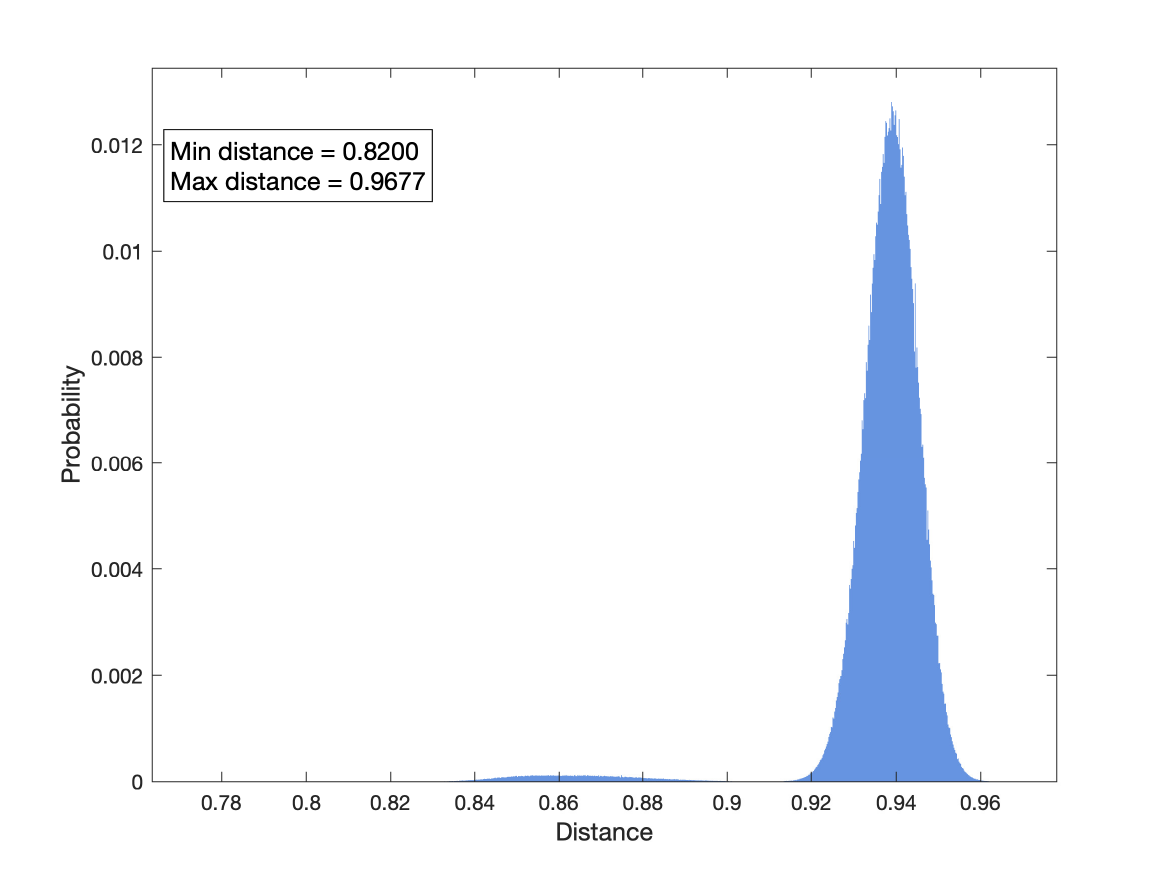}
  }
  \caption{Jaccard distance distributions for stochastic block models with given within block/between block connection probabilities.}
  \label{fig:sbm-graphs}
\end{figure}

As stated earlier, the goal of these comparisons is to validate the correct rejection of the assumption of similarity of the two graphs being compared. We conduct 12 tests between synthetic graphs with known (dissimilar) structures. Table~\ref{tab:pairwisecomp1} shows cases where the structural differences between the graphs are arguably slight. The distance distributions for each of the graphs referenced in this table are presented in Figure \ref{fig:er-graphs} (Erd\"os-R\'enyi, ER) and Figure \ref{fig:sbm-graphs} (stochastic block model, SBM). 

\begin{figure}[H]
  \centering
  \subfloat[CM3400]{%
    \includegraphics[width=0.48\textwidth]{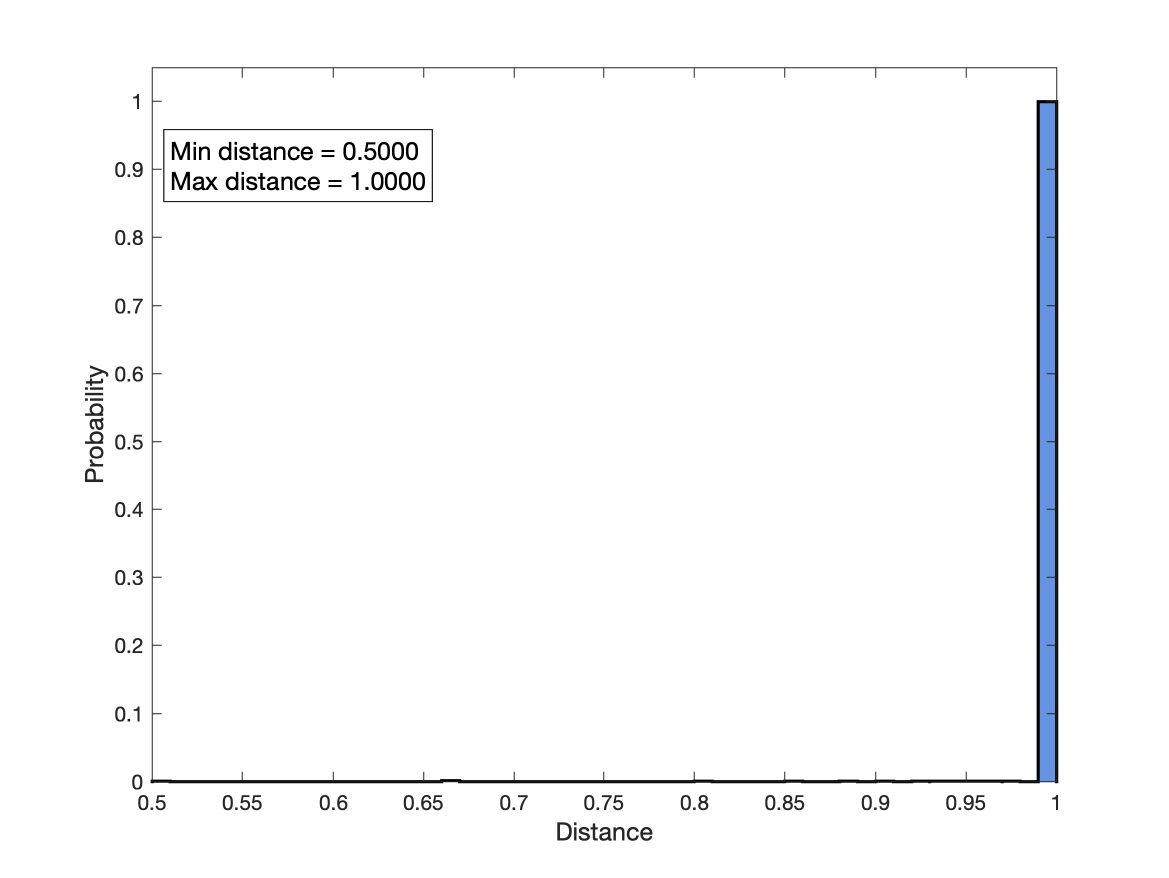}
  }\hfill
  \subfloat[CM3400 (Directed)]{%
    \includegraphics[width=0.48\textwidth]{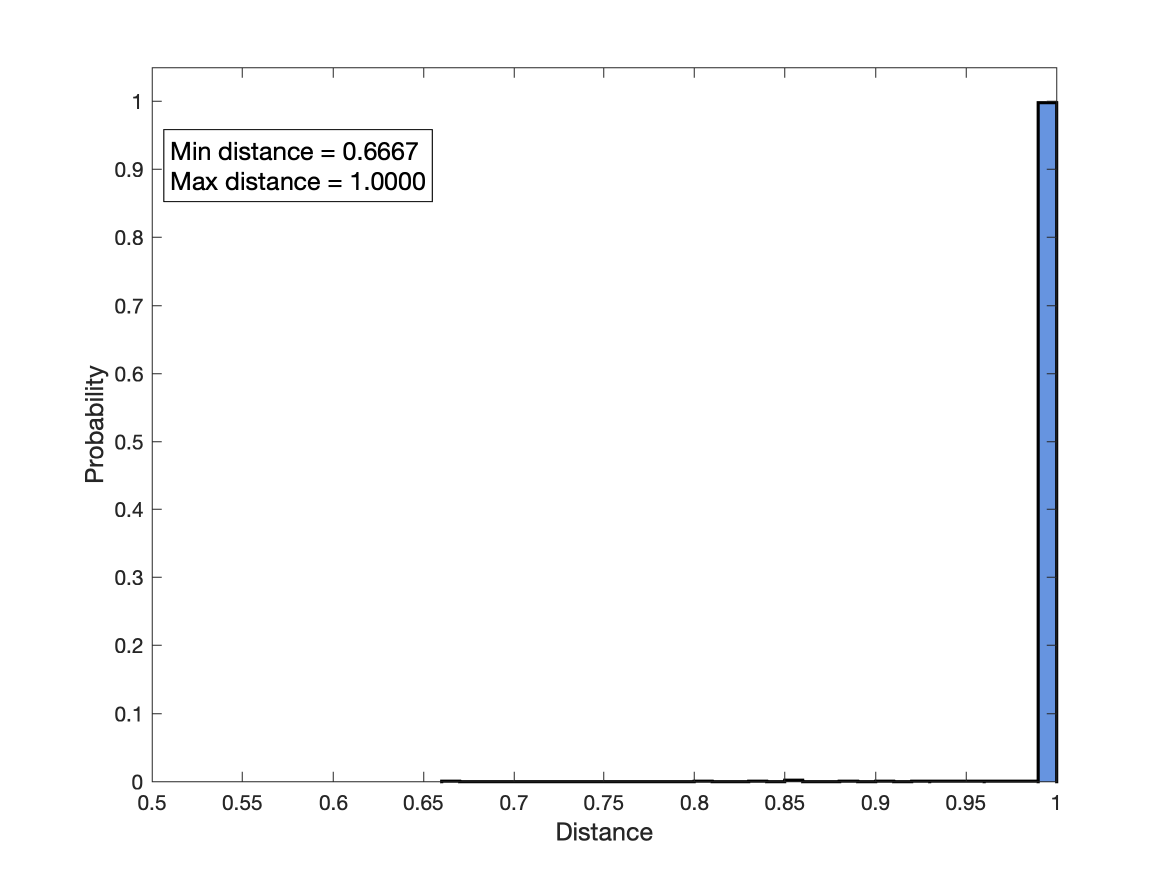}
  }
  \caption{Jaccard distance distributions for configuration models with 3400 nodes.}
  \label{fig:configuration-models}
\end{figure}

\begin{figure}[H]
  \centering
  \subfloat[CM vs ER 0.213]{%
    \includegraphics[width=0.48\textwidth]{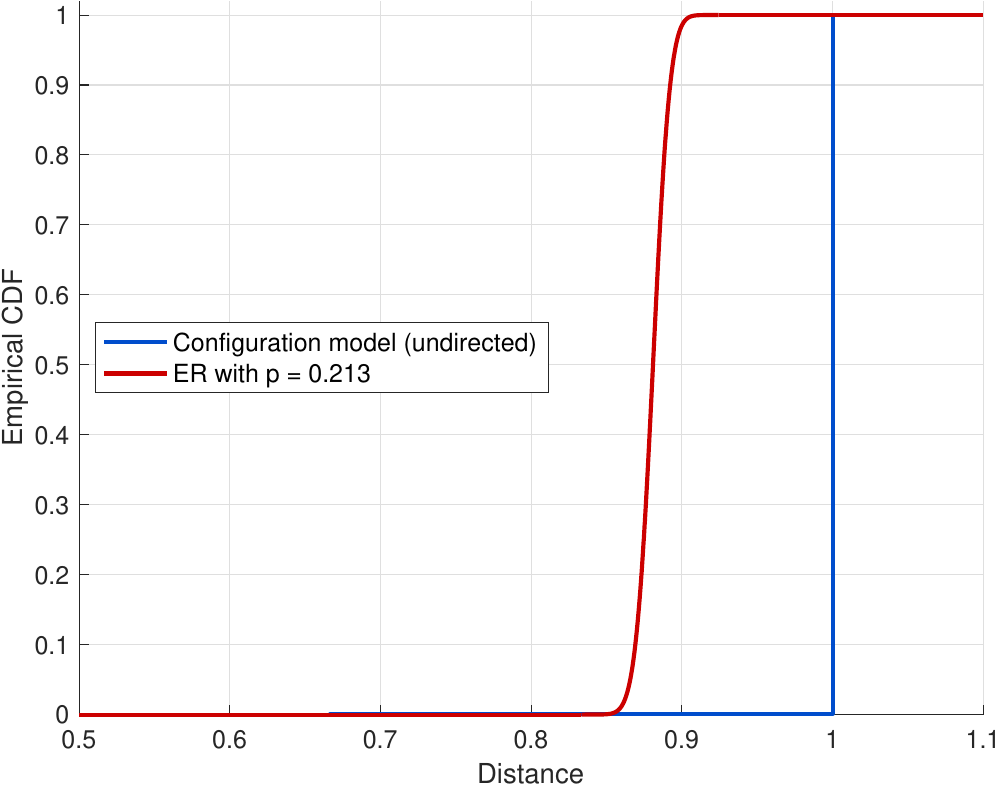}
  }\hfill
  \subfloat[CMD vs ER 0.213\ (Directed)]{%
    \includegraphics[width=0.48\textwidth]{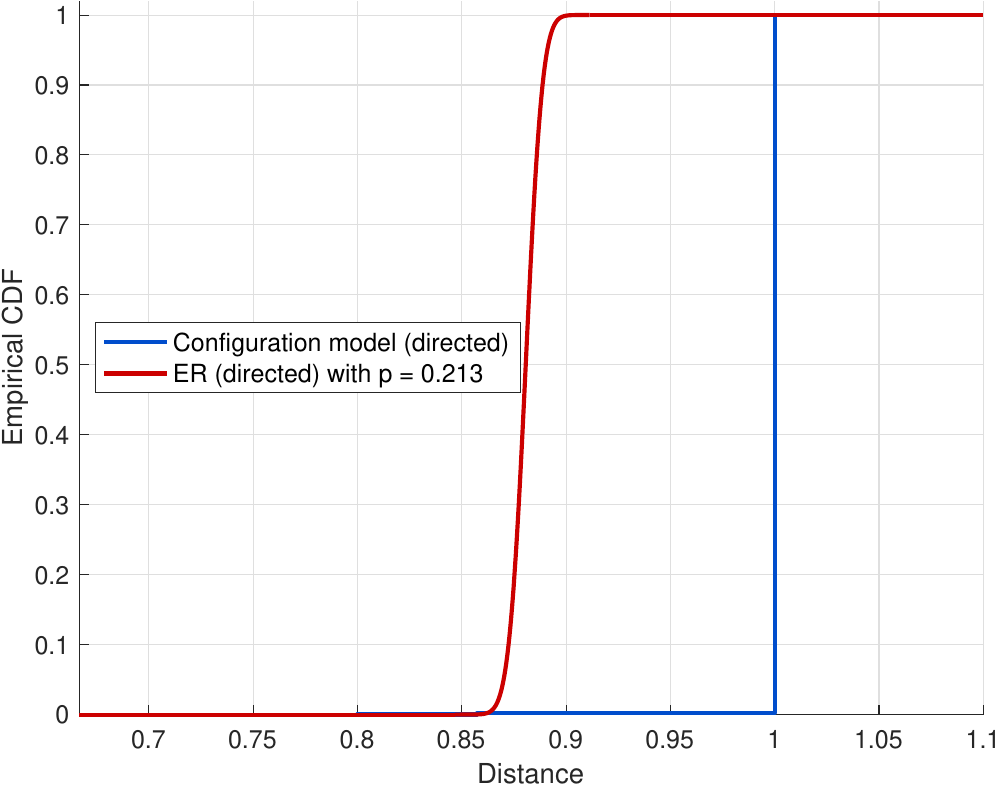}
  }\\[0.75ex]
  \subfloat[CM vs SBM 0.9/0.1]{%
    \includegraphics[width=0.48\textwidth]{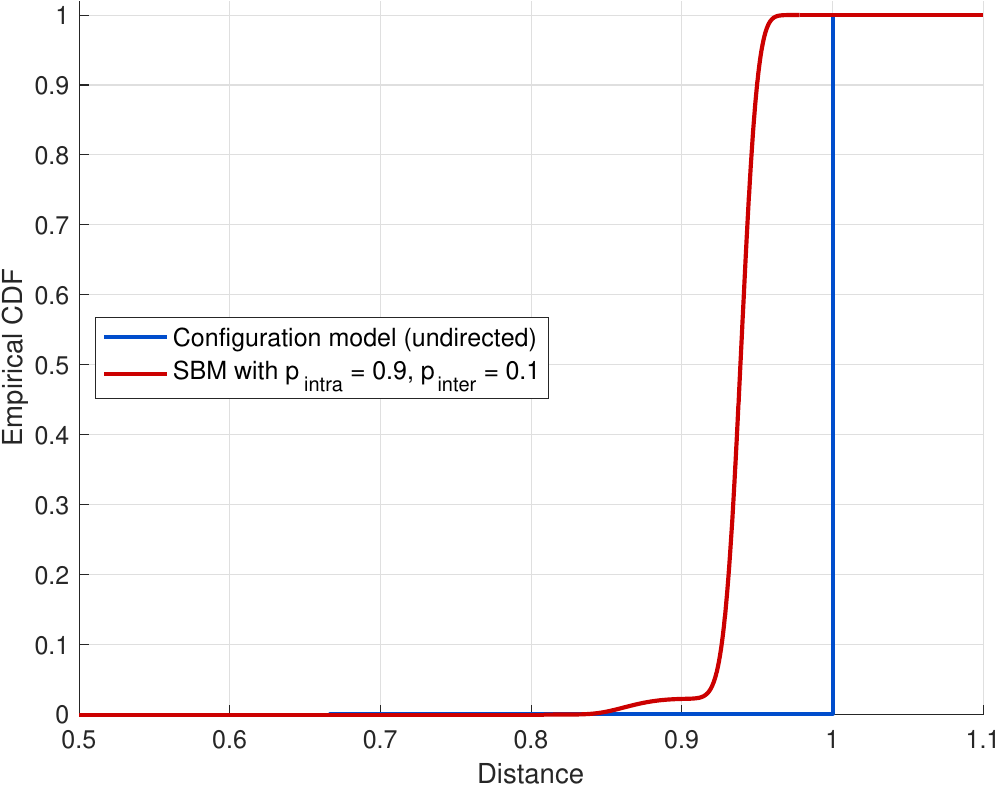}
  }\hfill
  \subfloat[CMD vs SBM 0.9/0.1\ (Directed)]{%
    \includegraphics[width=0.48\textwidth]{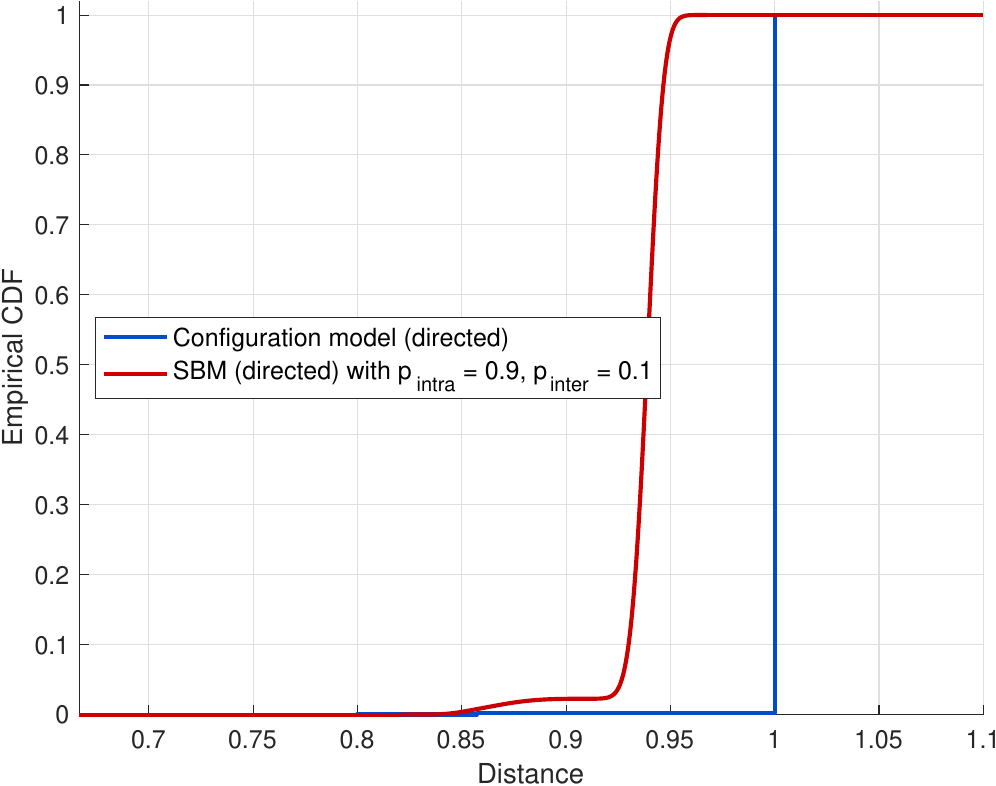}
  }\\[0.75ex]
  \subfloat[SBM 0.9/0.1 vs ER 0.213]{%
    \includegraphics[width=0.48\textwidth]{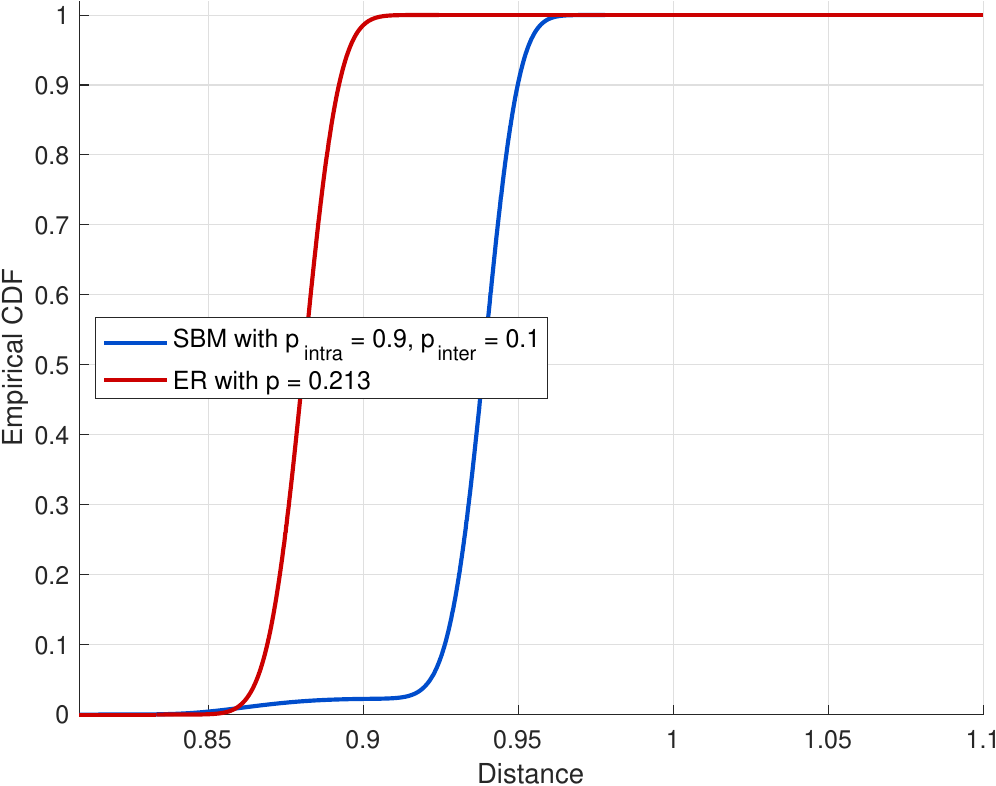}
  }\hfill
  \subfloat[SBM 0.9/0.1 vs ER 0.213\ (Directed)]{%
    \includegraphics[width=0.48\textwidth]{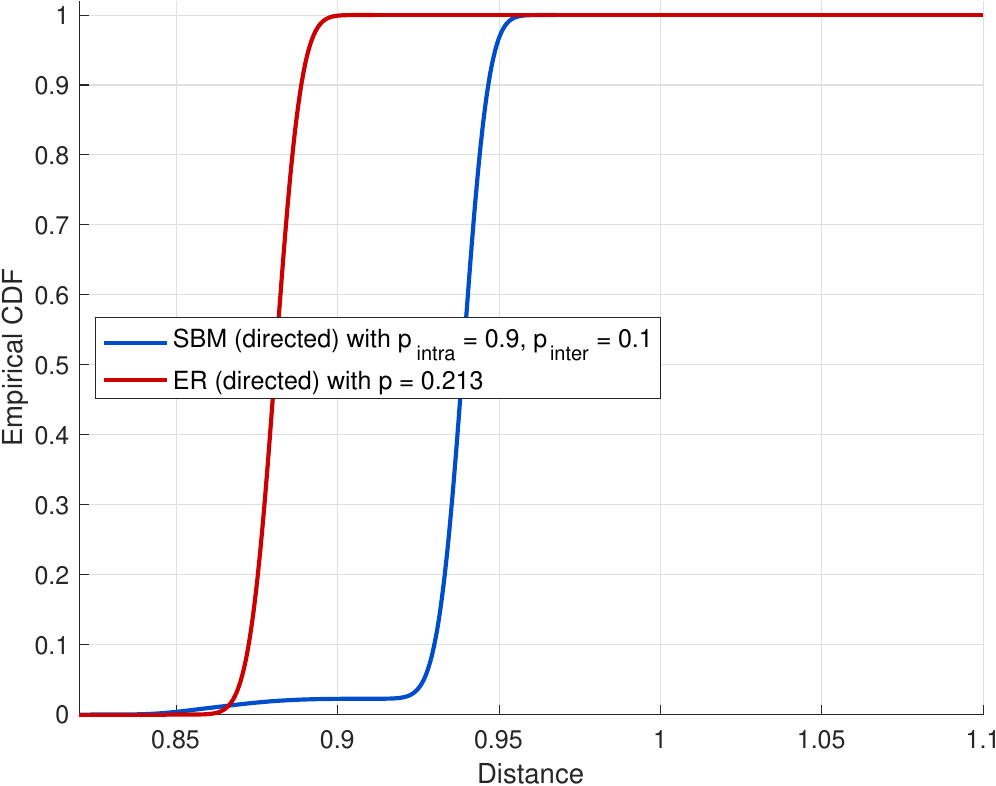}
  }
  \caption{Pairwise comparisons of CDFs of Jaccard distances.}
  \label{fig:distance-cdfs}
\end{figure}

Meanwhile, Table~\ref{tab:pairwisecomp2} displays tests results with graphs having marked structural differences. Distance distributions for the referenced configuration model (CM) models are presented in Figure \ref{fig:configuration-models}. The overlaid CDFs corresponding to each of the pairwise comparisons referenced in this table are presented in Figure \ref{fig:distance-cdfs}.

All synthetic graphs in these experiments have the same number of vertices ($N=3,400$). The SBM graphs are all composed of 45 blocks of varying sizes ($n \in [50,99]$) but uniform in/out edge probabilities. These probabilities are shown in their naming convention. The ER graphs against which they are compared were generated using edge probabilities that are equal to the density of the SBM graphs in the comparisons. The CM graphs were generated using power law in/out degree sequences of exponent 3.5.

% ===== Pairwise comparisons (1) =====
\begin{table}[H]
\centering
\tiny
\caption{Pairwise comparisons of ER graphs with given densities and SBM graphs with given within block/between block connection probabilities. (hard cases)}
\begin{tabular}{llllll}
\hline
Graph 1							& Graph 2					& KS Statistic 			\\
\hline
ER 0.309						& SBM 0.7/0.3				& 0.004 				 \\
ER 0.309 (Directed)				& SBM 0.7/0.3 (Directed)		& 0.005	      			 \\
ER 0.213							& SBM 0.8/0.2 			& 0.013				 \\
ER 0.213 (Directed)					& SBM 0.8/0.2 (Directed)		& 0.018			 \\
ER 0.118							& SBM 0.9/0.1				& 0.061				\\
ER 0.118 (Directed)					& SBM 0.9/0.1 (Directed)		& 0.086		\\
\hline
\end{tabular}
\label{tab:pairwisecomp1}
\end{table}

% ===== Pairwise comparisons (2) =====
\begin{table}[H]
\centering
\tiny
\caption{Pairwise comparisons of CM graphs with a given number of nodes (3400), ER graphs with given densities, and SBM graphs with given within block/between block connection probabilities. (easier cases)}  
\begin{tabular}{lllllll}
\hline
Graph 1				& Graph 2				& KS Statistic 			\\
\hline
CM3400				& ER 0.213			& 0.999	      			\\
CM3400 (Directed)		& ER 0.213 (Directed)	& 0.998			\\
CM3400				& SBM 0.9/0.1			& 0.999				\\
CM3400 (Directed)		& SBM 0.9/0.1 (Directed)	& 0.998 	\\
SBM 0.9/0.1 			& ER 0.213 			& 0.976				\\
SBM 0.9/0.1 (Directed)	& ER 0.213 (Directed)	& 0.977	 \\
\hline
\end{tabular}
\label{tab:pairwisecomp2}
\end{table}

% ===== Block 1: ER graphs (2x3 tile, undirected left / directed right) =====

% ===== Block 2: Stochastic block model graphs (2x3 tile, undirected left / directed right) =====

% ===== Block 3: Configuration models (2 side-by-side) =====

% ===== Block 4: Distance CDFs (2x3 tile, undirected left / directed right) =====

Our experiments clearly demonstrate the accuracy and value of the distribution of Jaccard distances as a summary of the graph and as a basis for comparisons between graphs.

\subsection{Sensitivity analysis} \label{sensitivity}
The tests in this section have three different goals. First we compare each graph to itself. Here, we expect the hypothesis of similarity not be rejected. We also compare each graph to a version of itself with a portion of its nodes or edges removed. The goal of these trials is to examine the test's sensitivity to various structural modifications. We compare every graph in this section to a version of itself with 0, 1, 5, 10, 25\% of its (randomly selected) nodes or edges removed. 

\subsubsection{Synthetic graphs}
We begin by removing a percentage of randomly selected nodes from each of our four ER graphs, and compare the distribution of Jaccard distances to the distribution of distances in the complete original graph. The K-S distances to the original graphs are presented in Table~\ref{tab:ernodes} through Table~\ref{tab:cmedges}.

% ===== ER-Nodes =====
\begin{table}[H]
\centering
\tiny
\caption{Nodes removal experiment for four ER graphs}
\begin{tabular}{lllllll}
\hline
Percent Removed 	&     0 &           1 &            5 &           10 &           25 \\
\hline
ER 0.5  		  	& 0 &  0.003 & 0.008 & 0.013 & 0.038 \\
ER 0.5 (Directed)  	& 0 &  0.002 & 0.007 & 0.018 &  0.039 \\
ER 0.333		  	& 0 &  0.003 & 0.008 & 0.016 & 0.038 \\
ER 0.333 (Directed) 	& 0 &  0.002 & 0.007 & 0.014 & 0.041 \\
\hline
\end{tabular}
\label{tab:ernodes}
\end{table}
We repeat the same experiments but remove a percentage of edges at random.
% ===== ER-Edges =====
\begin{table}[H]
\centering
\tiny
\caption{Edge removal experiment for four ER graphs}
\begin{tabular}{lllllll}
\hline
Percent Removed 	&     0 &           1 &            5 &           10 &           25 \\
\hline
ER 0.5  		  	& 0 & 0.189& 0.759 & 0.979 &   1 \\
ER 0.5 Directed  	& 0 & 0.263 & 0.902 & 0.999 &  1 \\
ER 0.333		  	& 0 & 0.104 & 0.481 & 0.799 &  0.998 \\
ER 0.333 Directed 	& 0 & 0.146 & 0.638 & 0.929 & 1 \\
\hline
\end{tabular}
\label{tab:eredges}
\end{table}
We then also repeat the same experiments (node and edge removal + distance comparison) on SBM graphs.
% ===== SBM-Nodes =====
\begin{table}[H]
\centering
\tiny
\caption{Nodes removal experiment for four SBM graphs}
\begin{tabular}{llllll}
\hline
Percent Removed 	&     0 &          1 &            5 &           10 &           25 \\
\hline
SBM 0.7/0.3  		& 0 & 0.003 &  0.006 & 0.014 & 0.041 \\
SBM 0.7/0.3 (Directed) & 0 & 0.002 & 0.007 & 0.013 & 0.042 \\
SBM 0.9/0.1 		 & 0 & 0.002 & 0.007 & 0.016 & 0.036 \\
SBM 0.9/0.1 (Directed) & 0 &  0.001 & 0.008 & 0.014 & 0.035 \\
\hline
\end{tabular}
\label{tab:sbmnodes}
\end{table}

% ===== SBM-Edges =====
\begin{table}[H]
\centering
\tiny
\caption{Edges removal experiment for four SBM graphs}
\begin{tabular}{llllll}
\hline
Percent Removed 	&     0 &        1 &            5 &           10 &           25 \\
\hline
SBM 0.7/0.3  		& 0 & 0.093 & 0.438 & 0.751 & 0.995 \\
SBM 0.7/0.3 (Directed) & 0 &  0.130 & 0.584 & 0.893 &  1 \\
SBM 0.9/0.1 		 & 0 & 0.029 & 0.146 & 0.284 & 0.628 \\
SBM 0.9/0.1 (Directed) & 0 & 0.041 & 0.202 & 0.389 & 0.785 \\
\hline
\end{tabular}
\label{tab:sbmedges}
\end{table}

We complete our tests on synthetic graphs with Configuration Model graphs.

% ===== CM-Nodes =====
\begin{table}[H]
\centering
\tiny
\caption{Nodes removal experiment for two CM graphs}
\begin{tabular}{llllll}
\hline
Percent Removed 	&     0 &          1 &            5 &           10 &           25 \\
\hline
CM  		& 0 & na &  na & na & na \\
CM (Directed) & 0 & 0.000 & 0.000 & 0.001 & na \\
\hline
\end{tabular}
\label{tab:cmnodes}
\end{table}

% ===== CM-Edges =====
\begin{table}[H]
\centering
\tiny
\caption{Edges removal experiment for two CM graphs}
\begin{tabular}{llllll}
\hline
Percent Removed 	&     0 &        1 &            5 &           10 &           25 \\
\hline
CM   		& 0 & 0.000 & na & na & na \\
CM (Directed) & 0 &  0.000 & 0.000 & na &  na \\
\hline
\end{tabular}
\label{tab:cmedges}
\end{table}
Here, we note that both CM graphs quickly become disconnected, as we remove nodes or edges, which renders our test inapplicable. In order to better understand why our distance does not seem to capture node or edge removals, in the cases were the graph remains connected, we examine the densities of these graphs, to understand if and how these edge and node removals affect the overall graph structure.

% ===== CM-Nodes-DENSITIES =====
\begin{table}[H]
\centering
\tiny
\caption{Densities under node removal experiments for two CM graphs}
\begin{tabular}{llllll}
\hline
Percent Removed 	&     0 &          1 &            5 &           10 &           25 \\
\hline
CM  		& 0.001 & na &  na & na & na \\
CM (Directed) & 0.001 & 0.001 & 0.001 & 0.001 & na \\
\hline
\end{tabular}
\label{tab:cmnodes}
\end{table}

% ===== CM-Edges-densities =====
\begin{table}[H]
\centering
\tiny
\caption{Densities under edge removal experiments for two CM graphs}
\begin{tabular}{llllll}
\hline
Percent Removed 	&     0 &        1 &            5 &           10 &           25 \\
\hline
CM   		& 0.001 & 0.001 & na & na & na \\
CM (Directed) & 0.001 &  0.001 & 0.001 & na &  na \\
\hline
\end{tabular}
\label{tab:cmedges}
\end{table}
Our experiments removed a percentage of existing edges or nodes. Naturally, removing a percentage of a very small number of pre-existing edges cannot significantly alter the graph. The removal of a few nodes or edges only has a very minor effect on this reality. Therefore, the graph's overall structure remains largely unaffected by our experiments and our test does not detect a meaningful change in the overall graph structure. Nevertheless, throughout the remainder of our experiments on synthetic graphs, our test produces results exactly as expected. The test detects dissimilarity as it increases and never classifies a graph as different from its unmodified self. We also note that, predictably, edge removal introduces a more marked variation in the graph than node removal.  

\subsubsection{Real-world graphs}
We also conduct computational tests on real-world graphs of varying sizes. We conducted our tests on the following (directed and undirected) graphs:
\begin{itemize}
\item email-Eu-core network \cite{Leskovec2007Graph,Yin2017local} (directed, largest connected component only, $N=986$)
\item Wikipedia vote network \cite{Leskovec2010Signed,Leskovec2010Predicting} (directed, largest connected component only, $N=7,066$)
\item Watts \& Strogatz power grid data \cite{WattsStrog1998} (undirected, $N=4,941$)
\item Facebook egonets \cite{Leskovec2012learn} (undirected, $N=4,039$)
\end{itemize}

% ===== Real-World Nodes =====
\begin{table}[H]
\centering
\tiny
\caption{Nodes removal experiment for real-world graphs}
\begin{tabular}{llllll}
\hline
Percent Removed 					&     0 &          1 &            5 &           10 &           25 \\
\hline
email-Eu-Core (Directed)  		  	& 0 &  0.002 & na & na & na \\
Wikipedia (Directed)  				& 0 &   na & na & na &  na \\
W \& S power grid		  				& 0 & na & na & na & na \\
Facebook egonets	 						& 0 &  0.000 & 0.003 & 0.007 & na\\
\hline
\end{tabular}
\label{tab:rwnodes}
\end{table}

% ===== Real-World Edges =====
\begin{table}[H]
\centering
\tiny
\caption{Edges removal experiment for real-world graphs}
\begin{tabular}{llllll}
\hline
Percent Removed 					&     0 &      1 &            5 &           10 &           25 \\
\hline
email-Eu-Core (Directed)  		  	& 0 &  na & na & na & na \\
Wikipedia (Directed)  				& 0 &  na & na & na &  na \\
W \& S power grid		  				& 0 & na & na & na & na \\
Facebook egonets	 						& 0 &  0.003 & na & na & na \\
\hline
\end{tabular}
\label{tab:rwedges}
\end{table}

Here too, the real world graphs are extremely sparse graph. The removal of a very small number of nodes or edges does not affect this reality. Therefore, the graph structure is largely unaffected. Our test results confirm this reality. Once again, to demonstrate how the node/edge removal in our experiments did not modify network structures, beyond causing disconnections, we show the evolution of each graph's density. In Table~\ref{tab:rwnodesDens} and Table~\ref{tab:rwedgesDens} we clearly observe that the node/edges removal have only minor effects on network structure, beyond disconnection (na).

% ===== Real-Nodes-densities =====
\begin{table}[H]
\centering
\tiny
\caption{Densities under node removal experiment for real-world graphs}
\begin{tabular}{llllll}
\hline
Percent Removed 					&     0 &          1 &            5 &           10 &           25 \\
\hline
email-Eu-Core (Directed)  		  	& 0.026 &  0.027 & na & na & na \\
Wikipedia (Directed)  				& 0.002 &   na & na & na &  na \\
W \& S power grid		  				& 0.001 & na & na & na & na \\
Facebook egonets	 						& 0.011 &  0.011 & 0.011 & na & na\\
\hline
\end{tabular}
\label{tab:rwnodesDens}
\end{table}

% ===== RW-Edges-densities =====
\begin{table}[H]
\centering
\tiny
\caption{Densities under edge removal experiment for real-world graphs}
\begin{tabular}{llllll}
\hline
Percent Removed 					&     0 &      1 &            5 &           10 &           25 \\
\hline
email-Eu-Core (Directed)  		  	& 0.026 &  na & na & na & na \\
Wikipedia (Directed)  				& 0.002 &  na & na & na &  na \\
W \& S power grid		  				& 0.001 & na & na & na & na \\
Facebook egonets	 						& 0.011 &  0.011 & na & na & na \\
\hline
\end{tabular}
\label{tab:rwedgesDens}
\end{table}

\subsubsection{Node and vertex additions} \label{adds}
In order to examine our test results on the network structure modification of very sparse graphs, we add nodes and random connections to other nodes on the graph and perform our test. For the experiments presented in Table~\ref{tab:addNodes}, we revisit the two synthetic (CM) and the four real-world graphs. These graphs were too sparse to be studied using node or edge removal, since they quickly, sometimes instantly, became disconnected.  Instead of removing edges or nodes, we add a pre-determined number of nodes to each graph. This number is a percentage of the number of pre-existing nodes. We also add the same percentage of connections to other randomly selected nodes in the graph.

\begin{table}[H]
\centering
\tiny
\caption{Node and edge addition}
\begin{tabular}{llllll}
\hline
Percent (of nodes and connections) Added 	&     0 &      1 &            5 &           10 &           25 \\
\hline
        CM 3400 (Directed) & 0 & 0.008 & 0.594 & 0.997 & 0.998 \\
        email-Eu-core (Directed) & 0 & 0.003 & 0.14 & 0.486 & 0.88 \\ 
        Wiki-Vote (Directed) & 0 & 0.014 & 0.752 & 0.916 & 0.986 \\ 
        CM 3400 & 0 & 0.004 & 0.374 & 0.972 & 0.999 \\ 
        W \& S power grid & 0 & 0.006 & 0.483 & 0.993 & 0.999 \\ 
        Facebook egonets & 0 & 0.006 & 0.363 & 0.811 & 0.956 \\ 
\hline
\end{tabular}
\label{tab:addNodes}
\end{table}

Finally, to contrast these experiments to node or edge removal ones, we show the evolution of graph densities in Table~\ref{tab:addNodesDens}. These final experiments highlight how our test will capture even a very small change in network structure. Along with all other experiments on the very sparse CM and real-world graphs, they also highlight how the significance of a change in structure is relative to the graph's initial state, not a universal scale, like a fixed percent change in the number of nodes or edges. 

\begin{table}[H]
\centering
\tiny
\caption{Densities under node and edge addition experiments}
\begin{tabular}{llllll}
\hline
Percent (of nodes and connections) Added 	&     0 &      1 &            5 &           10 &           25 \\
\hline
        CM 3400 (Directed) & 0.001 & 0.001 & 0.005 & 0.019 & 0.098 \\ 
        email-Eu-core (Directed) & 0.026 & 0.026 & 0.029 & 0.04 & 0.114 \\ 
        Wiki-Vote (Directed) & 0.002 & 0.002 & 0.007 & 0.02 & 0.099 \\ 
        CM 3400 & 0.001 & 0.001 & 0.005 & 0.019 & 0.098 \\
        Watts-Strogatz & 0.001 & 0.001 & 0.005 & 0.019 & 0.098 \\
        Fbook & 0.011 & 0.011 & 0.015 & 0.027 & 0.104 \\
\hline
\end{tabular}
\label{tab:addNodesDens}
\end{table}

\section{Discussion and conclusions}
We have shown that the distribution of Jaccard distances between its nodes provide a very accurate summary of a graph. We have also seen that the K-S distance between these distributions provides a very accurate estimation of the (dis)similarity between two graphs. These summarization and comparison properties have been demonstrated on a variety of directed and undirected, synthetic and real-world networks.

While our metric is unarguably accurate, we observe that in many cases the distance between networks may be small. For this reason, we suggest that our metric only be used in conjunction with domain expertise (e.g., knowledge of what is a typical graph variation) and a thorough examination of the graph's structure, when performing anomaly or change-point detection tasks. We believe that the significance of a distance is more a domain than a statistical question. In particular, our experiments with the very sparse (CM and real-world) graphs demonstrate that the significance in a change is relative to each graph's initial state, not an arbitrary variation in the number of edges or connections. 

Future work should focus on change-point and anomaly detection and on the validity of the $p$-values. Indeed, while we notice a clear monotonic decrease in $p$-values in step with increases in network dissimilarity, we would like to formally examine their statistical validity.

\section*{Acknowledgemens}
We thank Eglantine Camby and Gilles Caporossi of GERAD and HEC Montréal for their seminal work on the topic of node-node distances, which made this work possible.

\bibliography{biblio}

\end{document}